\documentclass[notoc,11pt]{JHEP}
\usepackage{amsmath}
\usepackage{amssymb}
\usepackage{amsfonts}
\usepackage{amscd}
\usepackage{bbm} 
\usepackage{fleqn} 
\usepackage[footnotesize]{caption}
\usepackage{graphicx} 
\usepackage{braket} 
\usepackage{mathrsfs}
\usepackage[center,footnotesize,hang]{subfigure}
\usepackage[bbgreekl]{MyBbol}
\usepackage{longtable}
\usepackage{accents}
\usepackage{BetterCite}
\newcommand{\CenterEps}[2][1]{
\ensuremath{\vcenter{\hbox{\includegraphics[scale=#1]{#2.eps}}}}
}% Input eps files: usage: \CenterEps[ScaleFactor]{FileName}

\newcommand{\RaiseBrace}[1]{\raise3pt\hbox{$\displaystyle#1$}}

\def\D{\mathrm{d}}

\def\<{\left\langle}
\def\>{\right\rangle}

\DeclareMathOperator{\diag}{diag}

\allowdisplaybreaks[3]
\preprint{TUM-HEP-475/02}

\title{%
Radiative Generation of
the LMA Solution from 
Small Solar Neutrino Mixing at the GUT Scale
}
\abstract{%
We show that in see-saw models with small 
or even vanishing lepton mixing angle $\theta_{12}$,
maximal $\theta_{23}$, zero $\theta_{13}$ and zero CP phases
at the GUT scale, 
the currently favored LMA solution of the solar neutrino problem 
can be obtained in a rather natural way by Renormalization Group effects.
We find that  
most of the running takes place in the energy ranges above 
and between the see-saw scales, 
unless the charged lepton Yukawa couplings are large, which would correspond to a
large $\tan \beta$ in the Minimal 
Supersymmetric Standard Model (MSSM). 
The Renormalization Group evolution of the 
solar mixing angle $\theta_{12}$ is generically 
larger than the evolution of $\theta_{13}$ and $\theta_{23}$. 
A large enhancement occurs for an inverted mass hierarchy and for a regular mass
hierarchy with $|m_2 - m_1| \ll |m_2 + m_1|$. 
We present numerical examples of the evolution of the lepton mixing angles
in the Standard Model and 
the MSSM, 
in which the current best-fit values of the LMA mixing angles are produced
with vanishing solar mixing angle $\theta_{12}$ at the GUT scale.}

\author{Stefan Antusch\\
Physik-Department T30, 
Technische Universit\"{a}t M\"{u}nchen\\ 
James-Franck-Stra{\ss}e,
85748 Garching, Germany\\
E-mail: \email{santusch@ph.tum.de} 
}
\author{Michael Ratz\\
Physik-Department T30, 
Technische Universit\"{a}t M\"{u}nchen\\ 
James-Franck-Stra{\ss}e,
85748 Garching, Germany\\
E-mail: \email{mratz@ph.tum.de} 
}

\begin{document}
\section{Introduction}

To compare experimental results with predictions from models 
defined at a high energy scale,
it is essential to evolve the parameters of the models from high to low energies
by the Renormalization Group Equations (RGE's).
At present, in the lepton sector the LMA solution of the solar neutrino problem with a large 
but non-maximal value of the solar
mixing angle $\theta_{12}$ is strongly favored by the experiments 
\cite{Ahmad:2002jz,Ahmad:2002ka,Barger:2002iv,Bandyopadhyay:2002xj,Bahcall:2002hv,deHolanda:2002pp}.
On the other hand, the mixing angles in the quark sector are found to be small.
It is therefore interesting to investigate if this discrepancy 
can be explained by the RG evolution of the lepton mixing angles,
i.e.\ by increasing them via RG running.

The possibility of increasing a 
small atmospheric mixing angle $\theta_{23}$ by RG effects
has been considered in \cite{Tanimoto:1995bf,Balaji:2000au}. 
The running of the solar mixing angle \(\theta_{12}\), 
starting at the mass scale of the heavy sterile neutrinos, 
was investigated in \cite{Miura:2000bj,Miura:2002nz}. 
Other authors who studied the 3 neutrino case
focused on nearly degenerate neutrinos
\cite{Ellis:1999my,Casas:1999tp,Casas:1999ac,Chankowski:2000fp,Chen:2001gk,Parida:2002gz},
on the existence of fixed points \cite{Chankowski:1999xc,Dutta:2002nq}, 
or on the effect of Majorana phases on 
the RG evolution of mixing angles \cite{Haba:2000tx}.  

We consider the see-saw scenario \cite{Gell-Mann:1980vs,Yanagida:1980}, 
i.e.\ the Standard Model (SM) or the Minimal 
Supersymmetric extension of the SM (MSSM)
extended by 3 heavy neutrinos that are singlets under the SM gauge
groups and have large explicit (Majorana) masses with a non-degenerate
spectrum.
Due to this non-degeneracy, one has to use several effective
theories with the singlets partly integrated out when studying the
evolution of the effective mass matrix of the light neutrinos
\cite{King:2000hk,Antusch:2002rr}.
Below the lowest mass threshold the neutrino mass matrix is given
by the effective dimension 5 neutrino mass operator in the SM or MSSM. 
The relevant RGE's were derived in 
\cite{Antusch:2002rr,Chankowski:1993tx,Babu:1993qv,Antusch:2001ck,Antusch:2002vn,Antusch:2002ek}.
Note that in a realistic GUT model, there can be contributions to the
$\beta$-functions from additional
particles and couplings below the GUT scale. 

In a recent study \cite{Antusch:2002hy}, we showed that starting with bimaximal 
mixing at the GUT scale, the large solar mixing of the LMA solution can be 
explained as an effect of the RG evolution of the mixing angles. A key
observation was that with bimaximal mixing and vanishing CP phases, 
the solar mixing angle changes considerably, while the
evolution of the other angles is comparatively small. 

In this paper, we assume 
arbitrary solar mixing $\theta_{12}$, maximal atmospheric mixing
$\theta_{23}$ and vanishing $\theta_{13}$ at the GUT scale.
Especially interesting is the configuration with zero $\theta_{12}$,
which we examine in detail.
In this study, we restrict ourselves to the case of vanishing CP phases
and positive eigenvalues of the effective neutrino mass matrix at the GUT scale.
We calculate the RG running numerically in order to obtain the mixing angles 
at low energy and to compare them with the experimentally favored values.
Extending the analysis of \cite{Antusch:2002hy}, we find that the  
evolution of the solar mixing angle 
is much larger than the evolution of the other angles.
Again, most of the running of the mixing angles takes place between 
and above the see-saw scales, which shows the importance of carefully 
studying the RG behavior in these regions.
We derive analytic formulae, which help to understand this effect.
The LMA solution of the solar neutrino problem 
can thus be obtained in a natural way from small or even vanishing solar mixing
at the GUT scale. 

\section{Solving the RGE's} 

To evolve the lepton mixing angles and neutrino
masses from the GUT scale to the electroweak (EW) or
Supersymmetry (SUSY)-breaking scale in a see-saw model, 
a series of effective theories has to be
used. These are obtained by successively integrating out the heavy singlets
at their mass thresholds, which are non-degenerate in general.
The derivation of the RGE's
and the method for dealing with these effective theories 
are given in \cite{Antusch:2002rr}. Starting at
the GUT scale, the strategy is to solve 
the systems
of coupled differential equations of the form
\begin{eqnarray}
\mu \frac{\D}{\D \mu}   \accentset{(n)}{X}_i
  = \accentset{(n)}{\beta}_{{X}_i} \RaiseBrace{\Bigl(}\RaiseBrace{\Bigl\{}
  \accentset{(n)}{X}_j\RaiseBrace{\Bigl\}}\RaiseBrace{\Bigl)}
\end{eqnarray}
for all the parameters $\accentset{(n)}{X}_i \in \RaiseBrace{\bigl\{}\accentset{(n)}{\kappa},\accentset{(n)}{Y_\nu},\accentset{(n)}{M},
\dots\RaiseBrace{\bigr\}}$ of the theory
in the energy ranges corresponding to the effective theories denoted by
$(n)$.
At each see-saw scale, tree-level matching is performed.
Due to the complicated structure of the set of differential equations, 
the exact solution can only be obtained numerically.
However, to understand certain features of the RG evolution,
an analytic approximation at the GUT scale will be derived in section
\ref{sec:AnalyticApproximation}.   

\subsection{Initial Conditions at the GUT Scale}

At the GUT scale $M_\mathrm{GUT}$, we assume maximal atmospheric 
mixing $\theta_{23}$ and vanishing $\theta_{13}$. 
We restrict ourselves to the case of positive mass eigenvalues and real
parameters, so that there is no CP violation. 
In the basis where the charged lepton Yukawa matrix
is diagonal, up to phase conventions the 
effective Majorana mass matrix of the light neutrinos is given by
\begin{eqnarray} \label{eq:BimaxFormofMnu}
 m_\nu|_{M_\mathrm{GUT}}
 = 
 V(\theta_{12}, 0, \tfrac{\pi}{4}) \cdot 
 m_\mathrm{diag} |_{M_\mathrm{GUT}} 
 \cdot V^T(\theta_{12}, 0, \tfrac{\pi}{4})\; ,
\end{eqnarray}
with
$m_\mathrm{diag} |_{M_\mathrm{GUT}} 
:= \diag\left(m_1
|_{M_\mathrm{GUT}},m_2|_{M_\mathrm{GUT}},m_3|_{M_\mathrm{GUT}}\right) $
and where
\begin{equation}
 V(\theta_{12},\theta_{13},\theta_{23})=\left(
 \begin{array}{ccc}
 c_{12}c_{13} & s_{12}c_{13} & s_{13}\\
 -c_{23}s_{12}-s_{23}s_{13}c_{12} &
 c_{23}c_{12}-s_{23}s_{13}s_{12} & s_{23}c_{13}\\
 s_{23}s_{12}-c_{23}s_{13}c_{12}&
 -s_{23}c_{12}-c_{23}s_{13}s_{12} & c_{23}c_{13}
 \end{array}
 \right)
\end{equation}
with \(s_{ij}=\sin\theta_{ij}\) and \(c_{ij}=\cos\theta_{ij}\)
is the (orthogonal) MNS matrix \cite{Maki:mu} 
in standard parametrization.
In our see-saw scenario, the effective mass matrix of the light
neutrinos is
\begin{equation}\label{eq:SeeSawF}
	m_\nu =
	\frac{v_\mathrm{EW}^2}{2} \, Y_\nu^T 
	M^{-1} Y_\nu
\end{equation}
at the high-energy scale,
with $\<\phi\> = \frac{v_\mathrm{EW}}{\sqrt{2}} \approx 174$ GeV.
Obviously, the neutrino Yukawa matrix  $Y_\nu$ and the singlet mass matrix $M$ 
cannot be determined uniquely from this relation, i.e.\ there is a set of
$\{Y_\nu, M\}$ configurations that yield single maximal mixing.
After choosing an initial condition for $Y_\nu$,
$M$ is fixed by the see-saw formula (\ref{eq:SeeSawF})
if $Y_\nu$ is invertible.
This determines the see-saw scales and thus the ranges of the various 
effective theories.

\subsection{Analytic Calculations} \label{sec:AnalyticApproximation}

The RGE for $m_\nu$ above the largest see-saw scale is given by
\begin{eqnarray} \label{eq:RGEforGamma}
  16\pi^2 \, \mu \frac{\D}{\D \mu} m_\nu
  & = & 
  C_e \left[Y_e^\dagger Y_e\right]^T m_\nu +
  C_e\,m_\nu \left[Y_e^\dagger Y_e\right] 
 \nonumber \\*
  &&{}+ 
  C_\nu \left[Y_\nu^\dagger Y_\nu\right]^T m_\nu +
  C_\nu\,m_\nu \left[Y_\nu^\dagger Y_\nu\right]
 \nonumber \\*
  &&{}+ \;\mbox{terms with trivial flavour structure}
  \label{eq:RunningOfCompositeMassOperator}
 \end{eqnarray}
 with $C_e = -\tfrac{3}{2}$, $C_\nu = \tfrac{1}{2}$ in the SM and
 $C_e = C_\nu = 1$ in the MSSM. The terms with trivial flavor-structure do not 
 influence the evolution of the mixing angles. 
 To calculate the ratios of the RG evolution 
 of the mixing angles at the
GUT scale, we further use the parametrization
 \begin{equation}\label{eq:mParamatrization}
  m_\nu(t)
  = 
  V\big(\theta_{12}(t),\theta_{13}(t),\theta_{23}(t)\big)
  \cdot m_\mathrm{diag}(t)\cdot 
  V^T\big(\theta_{12}(t),\theta_{13}(t),\theta_{23}(t)\big) \;,
 \end{equation}
where $\mu$ is the renormalization scale, $t := \ln \frac{\mu}{\mu_0}$,
and $m_\mathrm{diag} := \diag(m_1,m_2,m_3)$.
As in \cite{Antusch:2002hy}, we can parametrize the neutrino Yukawa coupling \(Y_\nu\) by
\begin{equation}\label{eq:GeneralParametrizationOfYnu}
  Y_\nu(y_1,y_2,y_3,\phi_{12},\phi_{13},\phi_{32}) 
  =
  \diag(y_1,y_2,y_3)  \cdot
  V^T(\phi_{12},\phi_{13},\phi_{32})\;.
\end{equation}
By differentiating equation \eqref{eq:mParamatrization} 
w.r.t.\ \(t\) and inserting the RGE
\eqref{eq:RGEforGamma}, we obtain analytic expressions for
\(\Dot\theta_{ij}\) and \(\Dot m_i\) at \(M_\mathrm{GUT}\), from which we can
compute the rather lengthy 
expression for \(\Dot\theta_{12}/\Dot\theta_{13}\)
and \(\Dot\theta_{12}/\Dot\theta_{23}\) at $M_\mathrm{GUT}$ 
as functions of the initial conditions.
Inserting the initial condition \(\theta_{13}=0\) and \(\theta_{23}=\frac{\pi}{4}\),
we find that generically the RG evolution of $\theta_{12}$ is much 
larger than the change of the other angles, unless $m_1$ is very small.  

For the special case \(\theta_{12}|_{M_\mathrm{GUT}}=0\), 
from these general formulae we obtain 
 \begin{subequations}\label{eq:AnalyticExpressions1} 
 \begin{eqnarray}
 \left.\frac{\Dot{\theta}_{12}}{\Dot{\theta}_{13}} \right|_{M_\mathrm{GUT}}
 & = &
 \frac{\left( m_{2} + m_{1} \right) \,
    \left( m_{3} - m_{1} \right) \,G_1}
	   {\left( m_{2} -  m_{1} \right) \,
    \left( m_{3} + m_{1} \right) \,G_2}
 \nonumber \\*
  & \approx &
  \left\{\begin{array}{ll}
   \displaystyle \pm \,
   \frac{m_2+m_1}{m_2-m_1} \,\frac{G_1}{G_2}&
   \mbox{for hierarchical neutrino masses\footnotemark[1]}
  \\*[3mm]
   \displaystyle \,
   \frac{\Delta m_\mathrm{atm}^2}{\Delta m_\mathrm{sol}^2} \,\frac{G_1}{G_2}&
   \mbox{for degenerate neutrino masses}
  \end{array}\right.
    \\*[2mm]
 \left.\frac{\Dot{\theta}_{12}}{\Dot{\theta}_{23}} \right|_{M_\mathrm{GUT}}
 & = &
 \frac{\left( m_{2} + m_{1} \right) \,
 	\left( m_{3} - m_{2} \right) G_1}
	{\left( m_{2} - m_{1} \right) \,
	 \left( m_{3} + m_{2} \right) \,G_3}\nonumber \\*
	 & \approx &
  \left\{\begin{array}{ll}
   \displaystyle \pm  \,
   \frac{m_2+m_1}{m_2-m_1} \,\frac{G_1}{G_3}&
   \mbox{for hierarchical neutrino masses\footnotemark[1]}
  \\*[3mm]
   \displaystyle \,
   \frac{\Delta m_\mathrm{atm}^2}{\Delta m_\mathrm{sol}^2} \,\frac{G_1}{G_3}&
   \mbox{for degenerate neutrino masses}
  \end{array}\right. 
\end{eqnarray}\footnotetext[1]{
	Note that this approximation is also valid for a relatively weak
	hierarchy, where $m_3$ is a few times larger or smaller than $m_1$, 
	$m_2$.
   }\end{subequations}   
where we neglected the contributions of $Y_e$ and 
\begin{subequations}\label{eq:AnalyticExpressions2}
\begin{eqnarray}
 G_1 
 & = &
 8\,\cos (\phi_{12})\,\left\{
 	\left( 2\,y_{1}^2 - y_{2}^2 - y_{3}^2 \right) 
		\,\cos (\phi_{13})\,
        \left[\sin (\phi_{13}) -\cos (\phi_{13})\, \sin (\phi_{12}) \right]
 \right\}\nonumber\\*
 & & {}
 	+ 4\,\cos (\phi_{12})\left\{\left( y_{2}^2 - y_{3}^2 \right) \,
          \cos (2\,\phi_{23})\,
          \left[ \left( 3 - \cos (2\,\phi_{13}) \right) 
		  	\,\sin (\phi_{12})  + 
            \sin (2\,\phi_{13}) \right]  \right\}
 \nonumber\\*  
 & & 
 {} + 8\,\left( y_{2}^2 - y_{3}^2 \right) \,
       \left( \cos (2\,\phi_{12})\,\sin (\phi_{13})
	   	-\cos (\phi_{13})\,\sin (\phi_{12})  \right) 
 \,\sin (2\,\phi_{23}) 
 \;,
 \\
 G_2 
 & = &
 {}-8\,\left( 2\,y_{1}^2 - y_{2}^2 - y_{3}^2 \right) 
 	\,\cos (\phi_{12})\,\cos (\phi_{13})\,
       \left( \cos (\phi_{13})\,\sin (\phi_{12}) + 
         \sin (\phi_{13}) \right)  
 \nonumber\\*
 & & {}- 4\,\left( y_{2}^2 - y_{3}^2 \right) \,
       \cos (\phi_{12})\,\cos (2\,\phi_{23})\,
       \left[ \left( \cos (2\,\phi_{13}) - 3 \right) \,
          \sin (\phi_{12}) + \sin (2\,\phi_{13}) \right]
 \nonumber\\*
 & &   
 {}+ 8\,\left( y_{2}^2 - y_{3}^2 \right) \,
       \left( \cos (\phi_{13})\,\sin (\phi_{12}) + 
         \cos (2\,\phi_{12})\,\sin (\phi_{13}) \right) 
 \,\sin (2\,\phi_{23})  
 \;,\\
 G_3
 & = &
 \sqrt{2}\, \left( 2\,y_{1}^2 - y_{2}^2 - y_{3}^2 \right) \,
 \left[ \cos^2 (\phi_{12}) + \left( \cos (2\,\phi_{12}) - 3 \right) \,
          \cos (2\,\phi_{13}) \right] 
 \nonumber\\*
 & & {}
 + \sqrt{2}\,\left( y_{2}^2 - y_{3}^2 \right) \,
       \left[\left( \cos (2\,\phi_{12}) - 3 \right) \, \cos (2\,\phi_{13}) 
	   		-6\,\cos^2 (\phi_{12}) 
	   \right] \,
       \cos (2\,\phi_{23})
 \nonumber\\*
 & & {} 
 + 4\sqrt{2}\,\left( y_{2}^2 - y_{3}^2 \right) \,
       \sin (2\,\phi_{12})\,\sin (\phi_{13})\,
       \sin (2\,\phi_{23}) 
 \;.
\end{eqnarray}
\end{subequations}
Note that the relation 
$ \Delta m_\mathrm{sol}^2\ll\Delta  m_\mathrm{atm}^2$
 also holds at the GUT scale.
Equations \eqref{eq:AnalyticExpressions1} and \eqref{eq:AnalyticExpressions2} 
can also be obtained from the formulae derived in
\cite{Casas:1999tg}.
The constants $G_1$, $G_2$ and $G_3$ clearly depend on the choice of
$Y_\nu|_{M_\mathrm{GUT}}$.  However, unless the parameters 
$\{y_1,y_2,y_3,\phi_{12},\phi_{13},\phi_{32}\}$ are fine-tuned, we
expect the ratios $G_1 / G_2$ and $G_1 / G_3$ to be of the order one. 
Thus, in the considered case of zero CP phases, 
the RG change of $\theta_{12}$ at the GUT scale is generically much larger than that of the
other angles, unless $m_1$ is very
small, in which case the ratio approaches 1.

\section{Numerical Examples with Vanishing $\theta_{12}$ at the GUT Scale}

In the previous section it was argued that the RG evolution of the solar mixing
angle is generically larger than the evolution of the other mixing angles.  
This raises the question whether the LMA
solution might be reached by RG evolution if one starts with 
vanishing $\theta_{12}$ and $\theta_{13}$ and 
maximal $\theta_{23}$ at \(M_\mathrm{GUT}\).  
An overview of the current allowed regions for the mixing angles 
and the mass squared differences is given in table \ref{tab:ExpData}. 
\begin{table}[!h]
\begin{center}
\vspace{-0.4cm}
\begin{longtable}{l|ccc}
& Best-fit value & Range (for \(\theta_{ij} \in [0^\circ,45^\circ]\)) & C.L. \\
  \hline
\(\theta_{12}\) [\({\:}^\circ\)]& \(32.9\) & \(26.1 - 43.3\) & 
\(99 \% 
\;(3\sigma)\)\\  
\(\theta_{23}\) [\({\:}^\circ\)]& \(45.0\) & \(33.2 - 45.0\) &    
\(99 \%
\;(3 \sigma)\)\\
\(\theta_{13}\) [\({\:}^\circ\)]& \(-\) &    
\(0.0-  9.2\) &     \(90 \%
\;(2 \sigma)\)\\
\(\Delta m^2_{\mathrm{sol}}\) [eV\(^2\)] & \(5 \cdot 10^{-5}\)&
\(2.3\cdot 10^{-5} - 3.7\cdot 10^{-4}\) & \(99 \%
\;(3 \sigma)\) \\
\(|\Delta m^2_{\mathrm{atm}}|\) [eV\(^2\)] & \(2.5 \cdot 10^{-3}\)&
\(1.2\cdot 10^{-3} - 5\cdot 10^{-3}\) &	\(99 \%
\;(3 \sigma)\)\hfill
\end{longtable}
\setcounter{table}{0}
\caption{
 Fits to the experimental data for the neutrino mixing angles and mass squared
 differences. 
 For the solar angle $\theta_{12}$ and the solar mass squared
 difference, the LMA solution has been assumed.
 The results stem from the analysis \cite{Bahcall:2002hv} 
 of the recent SNO data 
 \cite{Ahmad:2002jz,Ahmad:2002ka}, the Super-Kamiokande atmospheric data
 \cite{Toshito:2001dk} and the CHOOZ experiment \cite{Apollonio:1999ae}.
}
\label{tab:ExpData}
\vspace{-0.85cm}
\end{center}
\end{table}

We will investigate this possibility by
numerical calculations in the following.
We choose the specific form 
\begin{eqnarray}\label{eq:YnuExample}
Y_\nu = X \cdot 
\left(
\begin{array}{ccc}
\varepsilon^2 &\varepsilon^3&0\\
\varepsilon^3&\varepsilon &0\\
0&0&1
\end{array}
\right) 
\end{eqnarray}
for the neutrino Yukawa coupling at the GUT scale as an example. 
This example serves as an illustration. Of course, 
in many other textures large RG effects occur,
however, discussing them is beyond the scope of this study.
With a given $Y_\nu$, $M$ can be calculated from $m_\nu|_{M_\mathrm{GUT}}$ 
by the see-saw formula (\ref{eq:SeeSawF}). 

\subsection{Examples for the Running of the Lepton Mixing Angles}
\begin{figure}[!h]
        \begin{center}              
		\CenterEps{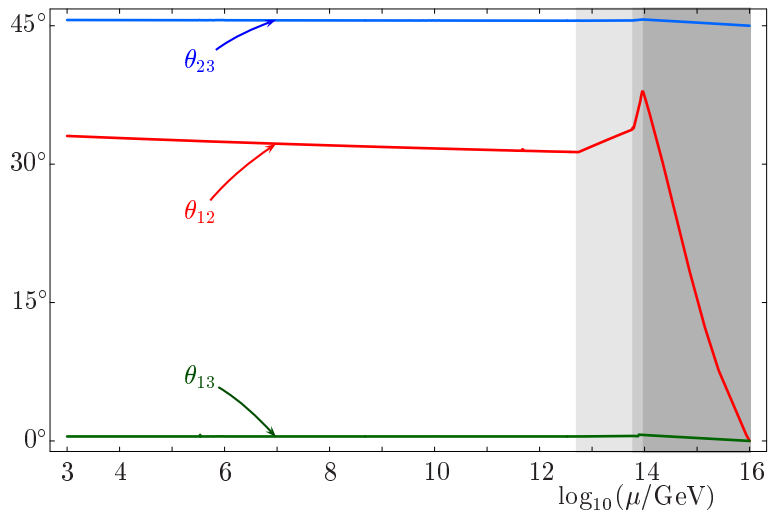}
        \end{center}
\caption{\label{fig:3} 
RG evolution of the mixing angles
from the GUT scale to the SUSY-breaking scale (taken to be $\approx 1$~TeV) 
in the MSSM extended by heavy singlets for a normal mass hierarchy
with $\tan\beta = 5$, $\varepsilon=0.65$, $m_1|_{M_\mathrm{GUT}}=0.076\,$eV and $X=0.5$.
In this example, the lightest neutrino has a mass of $0.05\,$eV. 
}
\end{figure}

\begin{figure}[!h]
       \begin{center}                   
		\CenterEps{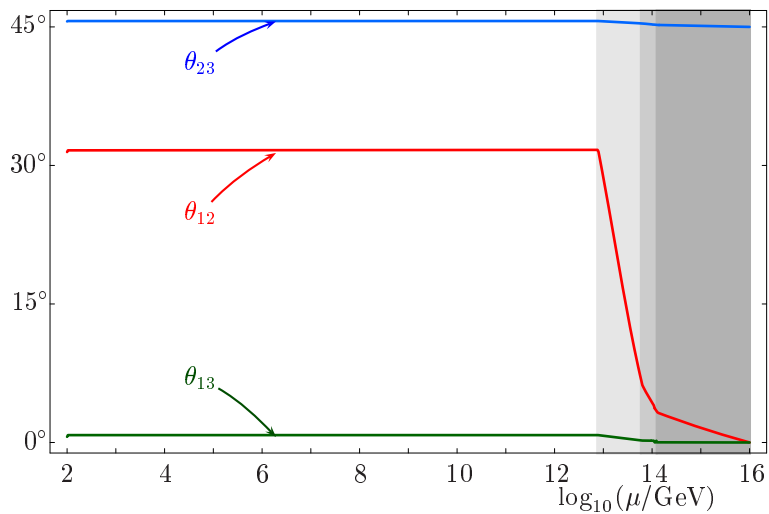}
        \end{center}
\caption{\label{fig:2} 
Example for the RG evolution of the mixing angles 
in the SM extended by heavy singlets 
from the GUT scale to the EW scale for a normal mass hierarchy
with $\varepsilon=0.6$, $m_1|_{M_\mathrm{GUT}}=0.049\,$eV and $X=0.5$. 
In this example, the lightest neutrino has a mass of $0.03\,$eV.
Most of the running takes place between the see-saw scales.   
}        
\end{figure}

Figures \ref{fig:3} and \ref{fig:2} show numerical 
examples for the
running of the lepton mixing angles from the GUT scale to the EW or
SUSY-breaking scale, which produce the experimentally favored mixing angles of
the LMA solution. 
The kinks in the plots 
correspond to the mass thresholds at the see-saw scales. 
The grey-shaded regions mark the various effective theories.

\subsection{Parameter Space Regions Compatible with the LMA Solution}
\label{sec:NumResults}

\subsubsection{Interpretation of the Parameters at the GUT Scale}

\begin{figure}[!h]
\begin{center}
\subfigure[Parameter $m_1 |_{M_\mathrm{GUT}}$
]{\label{subfig:Param-a}
\(\CenterEps[1]{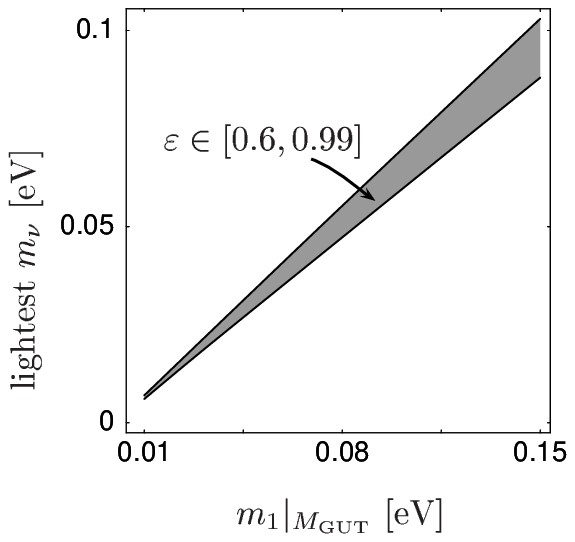}\)
}
\hfil
\subfigure[Parameter $\varepsilon$
]{\label{subfig:Param-e}
\(\CenterEps[1]{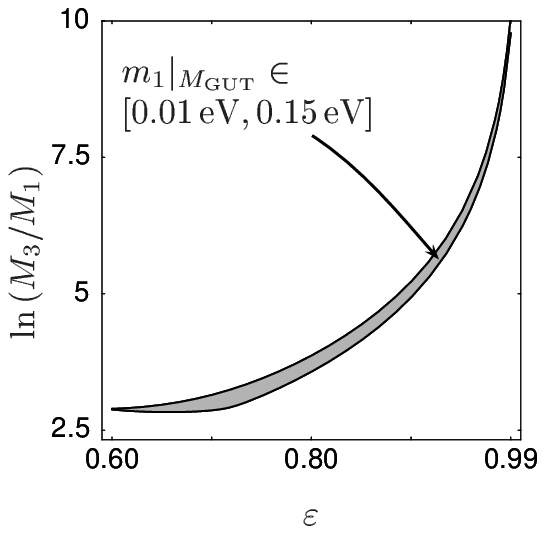}\)
}
\end{center}
\vspace*{-0.5cm}
\caption{
 Plot \ref{subfig:Param-a} shows the
 mass of the lightest neutrino (at low energy)
 as a function of $m_1|_{M_\mathrm{GUT}}$ for the SM and the MSSM with normal mass
 hierarchy, $X=0.5$ and $\varepsilon \in [0.6,0.99]$ (grey region). 
 Plot \ref{subfig:Param-e} shows the degeneracy of the see-saw scales,
 parametrized by $\ln (M_3/M_1)$ (at the GUT scale), as a
 function of $\varepsilon$ for the same cases with
 $m_1|_{M_\mathrm{GUT}} \in [0.01\text{ eV}, 0.15\text{ eV}]$ (grey region).
}
\label{fig:Parameters}
\end{figure}

The parameter $\varepsilon$ introduced in equation (\ref{eq:YnuExample})
controls the hierarchy of the entries in
$Y_\nu$ and thus the degeneracy of the see-saw scales.
Moreover, we choose the lightest neutrino mass at the GUT scale,
$m_1|_{M_\mathrm{GUT}}$ for a regular and $m_3|_{M_\mathrm{GUT}}$ for an inverted 
spectrum as a further initial condition.  
We fix the GUT scale values of the two remaining masses by the requirement
that the solar and atmospheric mass squared differences obtained at the
EW scale after the RG evolution be compatible with the allowed
experimental regions.  Thus, we are left with the free parameters $X$,
$\varepsilon$ and $m_1|_{M_\mathrm{GUT}}$ or $m_3|_{M_\mathrm{GUT}}$.
The dependence of the degeneracy of the see-saw scales on $\varepsilon$ and 
the mass of the lightest neutrino at the low scale from its initial value at the
GUT scale 
is shown in figure \ref{fig:Parameters}.
As mentioned above, we work in the basis where the Yukawa matrix of
the charged leptons is diagonal.

\subsubsection{Allowed Parameter Space Regions}
\begin{figure}[!h]
\begin{center}
\subfigure[SM
]{\label{subfig:SMr1Yinv}\(\CenterEps[1]{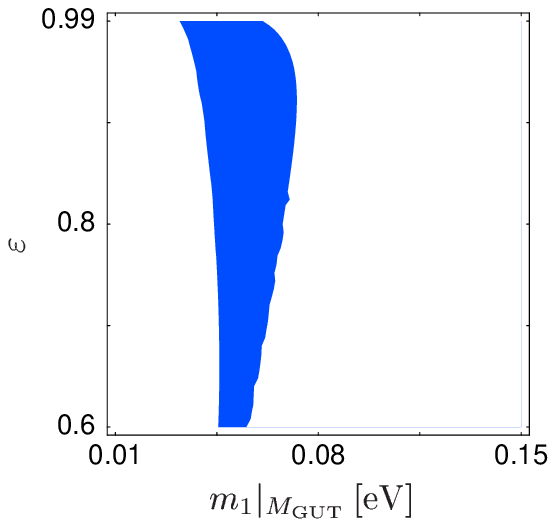}\)}
\hfil
\subfigure[MSSM
]{\label{subfig:MSSMr1Yinv}\(\CenterEps[1]{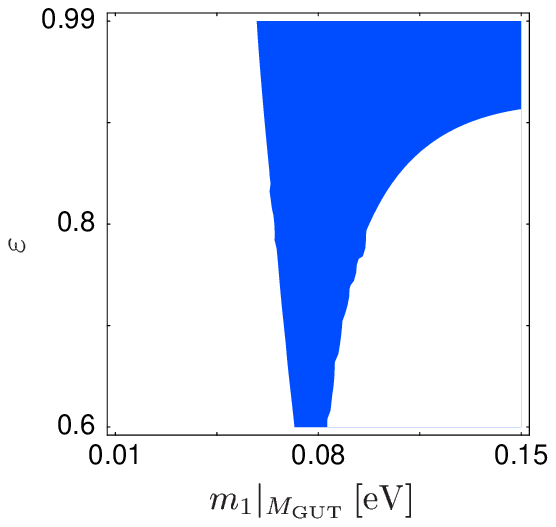}\)}
\end{center}
\vspace*{-0.5cm}
\caption{\label{fig:PScans} 
Parameter space regions compatible with the LMA solution of the solar neutrino
problem for the example with $Y_\nu$ given by equation (\ref{eq:YnuExample}).
The initial condition at the GUT scale $M_{\mathrm{GUT}}=10^{16}$ GeV is
vanishing mixing for $\theta_{12}$ and $\theta_{13}$ and maximal mixing for
$\theta_{23}$. The comparison with the experimental data is performed at
the EW scale or at 1~TeV for the SM 
and the MSSM, 
respectively.
The white regions of the plots are excluded by the data (LMA) at $3 \sigma$.
For this example, we consider the case of a normal neutrino mass hierarchy and
$X=0.5$ for the scale factor of the neutrino Yukawa couplings.
}
%\vspace{-0.5cm}
\end{figure}

The parameter space regions 
in which the RG evolution produces low-energy values
compatible with the LMA solution 
are shown in figure \ref{fig:PScans} for the SM and the MSSM ($\tan
\beta = 5$) with a normal mass hierarchy and $Y_\nu$ given 
in equation (\ref{eq:YnuExample}).   

We would like to stress that the shape of
the allowed parameter space regions strongly depends on the choice of 
the initial value of $Y_\nu$ at the GUT scale. 
For inverted neutrino mass spectra in the MSSM, allowed
parameter space regions exist as well. For the chosen range of 
initial conditions
at the GUT scale as in figure \ref{fig:PScans}, 
allowed regions with an inverted neutrino mass spectrum 
do not exist in the SM. 
One also has to ensure that the 
sign of $\Delta m^2_{\mathrm{sol}}$ is positive, as the LMA solution
requires this if the convention is used that the solar mixing angle is
smaller than $45^\circ$. 
For the chosen range of 
initial conditions there also exist regions, where the 
evolution of the solar
mixing angle is too large, i.e.\ the mixing angle 
would run above $45^\circ$.
This would correspond to a negative $\Delta m^2_{\mathrm{sol}}$ and thus these
regions have to be excluded.

\section{Summary and Conclusions}

We have studied the RG evolution of the lepton mixing angles 
in see-saw scenarios in the SM and in the MSSM. 
We have shown that the experimentally
favored neutrino mass parameters of the LMA solution 
can be obtained in a rather generic way from initial conditions with 
small or even zero solar mixing $\theta_{12}$, maximal
$\theta_{23}$ and 
 zero $\theta_{13}$ at the GUT scale.
 We have concentrated on the case of vanishing CP phases, which implies
positive mass eigenvalues\footnote{%
The general case of arbitrary CP phases is beyond the scope of this
study and will be investigated in a forthcoming paper \cite{Antusch:2002pr}.}.
We found that generically the RG evolution of the solar mixing angle 
is enhanced compared to the RG change of the two other mixings 
if one starts with maximal $\theta_{23}$, zero $\theta_{13}$ 
and arbitrary solar mixing $\theta_{12}$ at the GUT scale, and if 
the RG running is dominated by the neutrino Yukawa coupling.
A large enhancement occurs for an inverted mass hierarchy and for a regular mass
hierarchy with $|m_2 - m_1| \ll |m_2 + m_1|$.
For the special case of bimaximal GUT scale mixing, this has recently been observed in 
\cite{Antusch:2002hy}.
In the SM and MSSM with \(\theta_{12}\!=\!\theta_{13}\!=\!0^\circ\) 
and \(\theta_{23}\!=\!45^\circ\) as initial conditions at \(M_\mathrm{GUT}\), 
we have found regions in parameter space in which 
mixing angles compatible with the experimentally favored values of the LMA
solution are produced.
The effect does not require fine-tuning, 
degenerate neutrinos or a large value of $\tan \beta$
(in the case of the MSSM).
This opens up new possibilities for building models 
of neutrino masses.

\section*{Acknowledgements}
We would like to thank Manfred Lindner and J\"orn Kersten for 
useful discussions.
This work was supported in part by the 
``Sonderforschungsbereich~375 f\"ur Astro-Teilchenphysik der 
Deutschen Forschungsgemeinschaft''.
\goodbreak

\end{document}